\documentclass[prd,twocolumn,floats,floatfix,nofootinbib]{revtex4}
\usepackage[dvips]{graphicx}
\usepackage{graphics}
\usepackage{dcolumn}
\usepackage{amssymb}
\usepackage{bm}
\usepackage{amsmath}
\usepackage{color}
\usepackage{subfigure}
\usepackage{epstopdf}

\begin{document}

\title{Boosted Kerr-Newman Black Holes}

\author{Rafael F. Aranha\footnote{rafael.aranha@uerj.br}, Rodrigo Maier\footnote{rodrigo.maier@uerj.br}} 

\affiliation{
Departamento de F\'isica Te\'orica, Instituto de F\'isica, Universidade do Estado do Rio de Janeiro,\\
Rua S\~ao Francisco Xavier 524, Maracan\~a,\\
20550-900, Rio de Janeiro, Brazil
}

\date{\today}

\begin{abstract}
In this paper we obtain a new solution of Einstein field equations which describes a boosted Kerr-Newman black hole relative to a Lorentz frame at future null infinity. To simplify our analysis we consider a particular configuration in which the boost is aligned with the black hole angular momentum.
The boosted Kerr-Newman black hole is obtained considering the complete asymptotic Lorentz transformations of Robinson-Trautman coordinates to Bondi-Sachs, including the perturbation term of the boosted Robinson-Trautman metric.
To verify that the final form of the metric is indeed a solution of Einstein field
equations, we evaluate the corresponding energy-momentum tensor the boosted Kerr-Newman solution.
To this end, we consider the electromagnetic energy-momentum tensor
built with the Kerr boosted metric together with its timelike
killing vector. We show that the Papapetrou field thus obtained engender
an energy-momentum tensor which satisfies Einstein field equations up to 4th order
for the Kerr-Newman metric. To proceed, we examine the causal structure of the 
boosted Kerr-Newman black hole in Bondi-Sachs coordinates as in a preferred timelike foliation. We show that the
ultimate effect of a nonvanishing charge is to shrink the overall size of the event horizon and ergosphere areas when compared to the neutral boosted Kerr black holes. Considering the preferred timelike foliation we 
obtain the electromagnetic fields for a proper nonrotating frame of reference. We show that while the electric field displays a pure radial
behaviour, the magnetic counterpart develops an involved structure with 
two intense lobes of the magnetic field observed in the direction opposite to the boost.
\end{abstract}
\maketitle
\section{Introduction}
\label{intro}
In spite of the fact the physics of black holes is well established
in General Relativity -- at least from a theoretical point of view at classical level -- observations indicate that 
mass, angular momentum and charge do not furnish a
complete set of parameters to properly describe high energy configurations as remnants
black holes formed from binaries mergers. In fact, recent data by the LIGO and Virgo
collaborations\cite{virgo1,virgo2,virgo3,virgo4} established that gravitational wave emission from binary
black holes mergers were engendered from mass ratios ranging from $0.8$ down to
$0.5$ implying that the remnant black hole must have a boost along a particular
direction relative to an asymptotic Lorentz frame at null infinity – where such
emissions have been detected.

In the last decades several papers were devoted to the search of boosted black hole solutions. 
In this framework, numerical relativity has played a significant role to obtain initial data of spinning binary
black holes. In \cite{Karkowski:2006kp} initial data for boosted Kerr black holes were constructed
in the axially symmetric case. 
With the use of finite element method algorithms, momentum and hamiltonian constraints were
numerically solved and apparent horizons evaluated.
Kerr-Schild metrics\cite{kramer} of the form $g_{\mu\nu}=\eta_{\mu\nu}+h k_\mu k_\nu$
have also played an important role considering its invariance under Lorentz symmetry about
the Minkowski background $\eta_{\mu\nu}$. 
In fact, still in the context of numerical relativity
boosted Schwarzschild and Kerr spacetimes from their Kerr-Schild versions may
furnish initial data for black holes in a binary orbit\cite{Matzner:1998pt,Bonning:2003im,Cook:1997qc,Huq:2000qx}.
In \cite{Balasin:1995tj,Barrabes:2003up} the authors break the axial symmetry applying a Lorentz boost on Kerr-Schild 
coordinates of a Kerr
metric. All the above mentioned approaches are however rather different 
when compared to the results shown in \cite{Soares:2016bte,Soares:2020gqy,Aranha:2021zwf}. In this case the authors 
obtain an analytic approximate solution of a boosted Kerr black hole relative to a Lorentz frame at future
null infinity which should be expected to
correspond to the final configuration of a remnant black hole originated from 
the collision/merger of a system
of spinning black holes.

Apart from all the progress mentioned above, a final nonvanishing electric charge should complete the puzzle when
more general configurations take place. 
Although it is widely believed that charged black holes are rather unrealistic once they tend to attract opposite 
charges 
from their neighbourhood to neutralize themselves, a number of works have been developed in order to better 
understand 
proper mechanisms of charged black hole neutralization (see e.g.
\cite{Eardley:1975kp,Ruffini:2009hg,Hwang:2010im,Gong:2019aqa} and references therein).
As a direct astrophysical application it has been argued that 
binary black holes may admit electric charge. In fact, according to Zhang's mechanism\cite{Zhang:2016rli} a rotating
charged black hole may develop a magnetosphere which should allow a nonvanishing charge for a 
large period of time. The overall effect of such mechanism could provide a signature of an 
electromagnetic signal in gravitational wave events of binary black hole merger.
Further analyses have given support to electric charges in black holes considering 
electromagnetic counterparts of black hole mergers\cite{Liebling:2016orx,Liu:2016olx,Punsly:2016abn,Levin:2018mzg,Deng:2018wmy}.

From another observational point of view, the detection of charged black holes is characterized by the determination of their shadow \cite{bardeen,luminet}. In this context, we can initially highlight the work of Zakharov et al. \cite{zakharov}, where the authors considered the possibility of generalizing the parameters of black holes not only for the rotating case \cite{holz-wheeler} but also including the net electric charge in the shadow signature. Also according to Zakharov et al., the detection of the black hole shadow could be established by future space missions (at the time) such as RadioAtron in the radio band or MAXIM in the X-ray band.
Only in 2019, through the Event Horizon Telescope collaboration (EHT) \cite{akiyama2019}, the shadow of the black hole in M87* could be directly detected. In 2022, it was the time for the detection of the shadow of Sgr A* in the Milky Way \cite{akiyama2022}. The results are all consistent with the Kerr black hole solution, as predicted by General Relativity.
The EHT data delimits the observables of the black hole shadow, which provides a method of analyzing which theoretical model fits the results best. Especially by the fact that this method permits a comparison with other black hole solutions other than Kerr. This opens up the possibility for the detection of charged black hole solutions and also the delimitation of their charge values.

In a more recent context, we can highlight the work of Ghosh and Afrin \cite{ghosh-afrin}, where the authors impose an upper limit on the value of the black hole charge in Sgr A* using EHT data. It is important to say that this work not only considers limits for the charge, but also for parameters associated with more general black holes arising from modified theories of gravitation, such as Horndeski rotating black holes \cite{santiago-maier} and rotating hairy black holes \cite{contreras}. Thus, it is clear that new observational data on black hole shadows can be decisive in determining the types of black holes that exist in nature, both from a qualitative and quantitative point of view.

Within this context, in this work, we obtain a boosted Kerr-Newman black hole solution and analyze some of its basic features concerning the electric charge, the rotating and the boost parameters values. The method is described as follows. In his seminal paper\cite{Papapetrou:1966zz},  A. Papapetrou inferred that Killing vectors may be regarded as vector
potentials
which generate electromagnetic fields – Papapetrou fields – satisfying the covariant
Lorentz gauge and Maxwell-like equations.
One of the remarkable features of Papapetrou field is its potential
to lead to solutions of Einstein field equations which
describe physical configurations.
In fact, it can be shown\cite{Maier:2023xhl} that the physical electromagnetic field of
a Kerr-Newman black hole can be generated directly from the Papapetrou field obtained
from the timelike Killing vector of a Kerr spacetime. In the case of a vanishing rotation, the physical 
electromagnetic field of
a Reissner-Nordstr\"om black hole is the same as the Papapetrou field generated from a
timelike Killing vector of a Schwarzschild spacetime, as one should expect. Following
a similar approach one should expect that the Papapetrou fields generated by boosted
rotating black holes\cite{Aranha:2021zwf} could lead to boosted Kerr-Newman solutions. 
We explicitly show this mechanism in this paper.

We organize the paper as follows. In the next section we obtain a new solution of Einstein field equations which describes a boosted Kerr-Newman black hole relative to a Lorentz frame at future null infinity. In Section III we examine the resulting spacetime structure due to its event horizon
and ergosphere formation. In Section IV we evaluate the electromagnetic fields considering
a preferred timelike foliation for a locally nonrotating observer. In Section V we leave our final remarks.

\section{Boosted Kerr-Newman Black Hole}

In order to obtain the Kerr-Newman metric of a boosted black hole we will follow a standard procedure based on \cite{Aranha:2021zwf,Soares:2016bte,Soares:2020gqy}. 
We start by considering the Kerr-Newman metric in Robinson-Trautman coordinates $(u, r, \theta,\phi)$
\begin{eqnarray}
\nonumber
ds^2=\frac{r^2+\Sigma^2(\theta)}{K^2(\theta)}(d\theta^2+\sin^2\theta d\phi^2)~~~~~~~~~~~~~~~~~~\\
\label{igrg}
-2\Big[du+\frac{\omega_0 \sin^2\theta}{K^2(\theta)}d\phi\Big]\Big[dr-\frac{\omega_0\sin^2\theta}{K^2(\theta)}d\phi\Big]~~~~~~~~~\\
\nonumber
-\Big[du+\frac{\omega_0 \sin^2\theta}{K^2(\theta)}d\phi\Big]^2\Big[1-\frac{(2m_0 r-q_0^2)}{r^2+\Sigma^2(\theta)}\Big],
\end{eqnarray}
where
\begin{eqnarray}
\label{kth}
K(\theta)&=&a+b\cos \theta\\
\label{sig}
\Sigma(\theta)&=&\frac{\omega_0}{K(\theta)}(b+a\cos\theta),
\end{eqnarray}
with $a=\cosh\gamma$ and $b=\sinh\gamma$. In the above, $\omega_0$, $m_0$ and $q_0$ are respectively connected to the rotation, mass and charge parameters while $\gamma$ stands for the boost parameter.

The Robinson-Trautman coordinates are related to the Bondi-Sachs asymptotic configuration $(U, R, \theta, \phi)$ through the following transformations\cite{Aranha:2013rj} 
\begin{eqnarray}
\label{rit1}
r\simeq K(\theta) R,~~ dr\simeq K(\theta) dR, ~~du\simeq\frac{dU}{K(\theta)}.
\end{eqnarray}
Moreover, a careful analysis indicates\cite{Soares:2020gqy} that the rotation and mass parameters
should transform as 
\begin{eqnarray}
\label{tp}
&&\omega_0= \omega_b(\theta) K(\theta),\\
&&m_0=m_b(\theta)K^3(\theta),
\end{eqnarray}
while
\begin{eqnarray}
\label{qb}
q^2_0=q^2_b(\theta)K^3(\theta).
\end{eqnarray}
Using (\ref{rit1})--(\ref{qb}) in (\ref{igrg}) the boosted metric asymptotically assumes the form
\begin{eqnarray}
\nonumber
ds^2=(R^2+\tilde{\Sigma}^2)(d\theta^2 + \sin^2\theta d\phi^2)~~~~~~~~~~~~~~~~~~~~~~\\
\label{ireply}
-2(dU+\omega_b \sin^2\theta d\phi)\Big(dR-\frac{\omega_b\sin^2\theta}{K^2}d\phi\Big)~~~~~~~~~~~~\\
\nonumber
-(dU+\omega_b \sin^2\theta d\phi)^2\Big(\frac{1}{K^2}-\frac{(2m_b R-q_b^2/K)}{R^2+\tilde{\Sigma}^2}\Big)\\
\nonumber
+\mathcal{O}\Big(\frac{1}{R^2}\Big),
\end{eqnarray}
in the Bondi-Sachs coordinates. In the above we have defined $\tilde{\Sigma}\equiv\Sigma/K$.

At this stage it is opportune to obtain the boosted Kerr-Newman
metric (\ref{ireply}) in a preferred timelike foliation.
To this end, we apply (\ref{rit1}) together with $du=dt-dr$ so that (\ref{ireply}) 
can be written in the coordinates $(t, r, \theta, \phi)$ as 
\begin{widetext}
\begin{eqnarray}
\label{ks}
\nonumber
ds^2=-\Big[1-\frac{(2 m_br  -q_b^2)K^3}{r^2+{\Sigma}^2}\Big]dt^2
-\frac{2(2m_br-q_b^2)K^3}{r^2+{\Sigma}^2}dtdr
+\frac{2(2m_b r-q_b^2)\omega_bK^2 \sin^2{\theta}}{(r^2+{\Sigma}^2)}dtd\Phi~~~~~~~~~~\\
+\Big[1+\frac{(2m_b r  -q_b^2)K^3 }{r^2+{\Sigma}^2}\Big]dr^2
-\frac{2\omega_b\sin^2\theta}{K}\Big[1+\frac{(2m_br-q_b^2)K^3}{r^2+{\Sigma}^2}\Big]drd\phi
+\frac{r^2+{\Sigma}^2}{K^2}d\theta^2\\
\nonumber
+\frac{\sin^2\theta}{K^2}\Big\{r^2+{\Sigma}^2
+{\omega_b^2\sin^2\theta}\Big[1+\frac{(2m_b r -q_b^2)K^3}{r^2+{\Sigma}^2}         \Big]
\Big\}
d\Phi^2+ \mathcal{O}\Big(\frac{1}{r^2}\Big).
\end{eqnarray}
\end{widetext}
It can be easily shown that the nonvanishing components (up to 4th order in $1/r$) of the Einstein tensor then read
\begin{eqnarray}
&&G^t_{~t}\simeq-\frac{q_b^2K^3}{r^4},~~G^{t}_{~\phi}\simeq -\frac{2q_b^2\omega_b\sin^2\theta K^2}{r^4},\\
&&G^r_{~r}\simeq -\frac{q_b^2K^3}{r^4},~~G^\theta_{~\theta}\simeq
\frac{q_b^2K^3}{r^4},\\
&&G^\phi_{~\phi}\simeq\frac{q_b^2K^3}{r^4}.
\end{eqnarray}
To verify that (\ref{ks}) is indeed a solution of Einstein field equations up to 4th order in $1/r$,
we evaluate the energy-momentum tensor the boosted Kerr-Newman solution. To this end, let us consider the energy-momentum tensor
\begin{eqnarray}
\label{tmunu}
T^\mu_{~\nu}=F^{\mu\gamma}F_{\nu\gamma}-\frac{1}{4}\delta^{\mu}_{\nu}F^{\alpha\beta}F_{\alpha\beta}    
\end{eqnarray}
built with the Kerr boosted metric ((\ref{ks}) with $q_b=0$). Assuming that the $4$-vector
$A^\mu$ is given by the timelike killing vector of the Kerr boosted metric, namely, 
$ A^\mu=\epsilon_0\delta^\mu_{~t}$, where $\epsilon_0$ is a constant parameter,   
we obtain that the nonvanishing components (up to 4th order in $1/r$) of (\ref{tmunu})
read
\begin{eqnarray}
\label{t1}
&&T^t_{~t}\simeq-\frac{2m_b^2K^6\epsilon_0^2}{r^4},~~T^{t}_{~\phi}\simeq -\frac{4m_b^2\omega_b\sin^2\theta K^5\epsilon_0^2}{r^4},~~~~\\
\label{t2}
&&T^r_{~r}\simeq -\frac{2m_b^2K^6\epsilon_0^2}{r^4},~~T^\theta_{~\theta}\simeq
\frac{2m_b^2K^6\epsilon_0^2}{r^4},\\
\label{t4}
&&T^\phi_{~\phi}\simeq\frac{2m_b^2K^6\epsilon_0^2}{r^4}.
\end{eqnarray}
Therefore, fixing
\begin{eqnarray}
\label{t5}
\epsilon_0=\frac{1}{\sqrt{2}}\Big(\frac{q_0}{\kappa m_0}\Big)    
\end{eqnarray}
where $\kappa^2$ is the Einstein's constant, it is easy to see that 
\begin{eqnarray}
G^{\mu}_{~\nu}=\kappa^2 T^\mu_{~\nu},    
\end{eqnarray}
up to 4th order in $1/r$.

\section{Spacetime Structure}

In order to understand the spacetime structure of the metric which describes our
boosted Kerr-Newman black hole, we consider two fundamental surfaces, namely the event horizon and ergosphere.

To start, let us consider our solution (\ref{ireply}) in Bondi-Sachs coordinates.
In this case, coordinate singularities are given by the condition $g^{RR}=0$, or
\begin{eqnarray}
\label{eqh}
R^2-2m_bRK^2+\tilde{\Sigma}^2+\frac{\omega_b^2\sin^2\theta}{K^2}+q_b^2K=0.    
\end{eqnarray}
From (\ref{eqh}) one may obtain an event horizon $R_h$ given by
\begin{eqnarray}
R_h=m_bK^2+\sqrt{m_b^2K^4-\omega_b^2-q_b^2K}, 
\end{eqnarray}
as long as 
\begin{eqnarray}
\label{cch1}
m_b^2K^4\geq\omega_b^2+q_b^2K.    
\end{eqnarray}
From (\ref{cch1}) we see that there is an upper limit for that charge parameter
-- once the mass and angular momentum are fixed -- which allows the formation
of an event horizon thus satisfying the cosmic censorship hypothesis\cite{Penrose:1969pc}.  
\begin{figure*}
\begin{center}
\includegraphics[width=5cm,height=5cm]{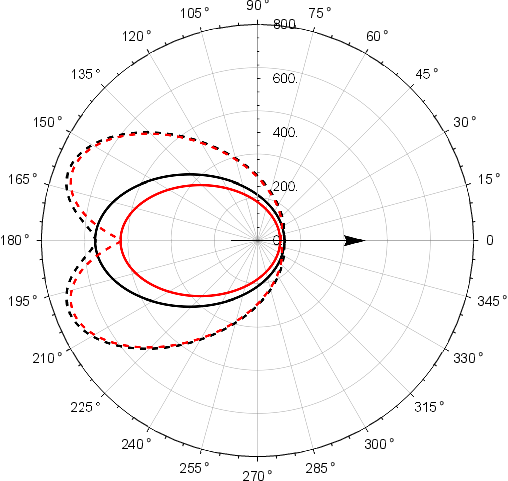}
\includegraphics[width=5cm,height=5cm]{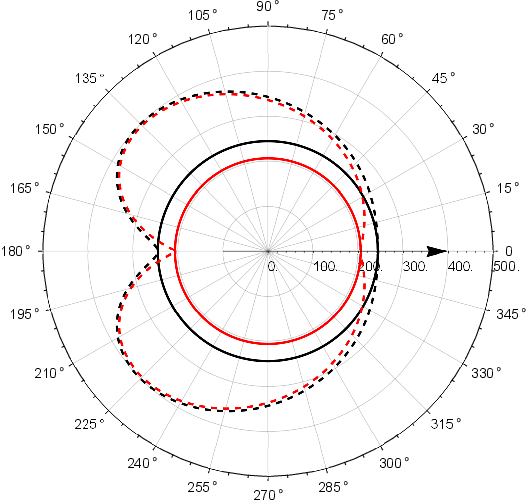}
\caption{The event horizons and ergospheres of boosted black holes -- solid and dashed curves, respectively. Thin arrows indicate the angular momentum/boost directions. Black (red) curves are connected boosted Kerr (Kerr-Newman) black holes. In the left panel we are using Bondi-Sachs coordinates
so that there is an explicit deformation of the event horizon/ergosphere in the 
boost direction. In the right panel we considered the preferred timelike foliation in which the metric is written in the
$(t, r, \theta, \phi)$ coordinates. Here we see that the event horizon is spherically symmetric while the ergosphere is deformed in the boost direction. In both panels we fixed the parameters 
$m_0=200$, $\omega_0=195$, $q_0=44$ and $\gamma=0.9$.
Here we see that a nonvanishing electric charge
diminishes the overall domain of the event horizon and ergosphere.}
\label{fig1}
\end{center}
\end{figure*}

The ergosphere, on the other hand, is defined by the condition $g_{UU}=0$, or
\begin{eqnarray}
\label{erg1}
R^2-2m_bRK^2+\tilde{\Sigma}^2+q_b^2K=0.    
\end{eqnarray}
From (\ref{erg1}) one may obtain the ergosphere radius $R_e$ as
\begin{eqnarray}
R_e=m_bK^2+\sqrt{m_b^2K^4-\tilde{\Sigma}^2-q_b^2K},   
\end{eqnarray}
as long as
\begin{eqnarray}
\label{cch}
m_b^2K^4\geq\tilde{\Sigma}^2+q_b^2K.    
\end{eqnarray}

To proceed we now consider the event horizon and ergosphere surfaces in the preferred timelike foliation used in (\ref{ks}). In this case we obtain
\begin{eqnarray}
g^{rr}=1+\frac{\omega_b^2\sin^2\theta-(2m_b r-q_b^2)K^3}{r^2+{\Sigma}^2}   
\end{eqnarray}
so that the horizon condition $g^{rr}=0$ furnishes a spherically symmetric event horizon
\begin{eqnarray}
r_{h}=m_0+\sqrt{m_0-\omega_0^2-q_0^2}.    
\end{eqnarray}
The ergosphere radius $r_e$, on the other hand, is defined by $g_{tt}=0$.
According to (\ref{ks}) we then obtain
\begin{eqnarray}
\label{ergt}
r_{e}=m_0+\sqrt{m^2_0-{\Sigma}^2-q^2_0}.    
\end{eqnarray}

Comparing these results with the boosted Kerr case\cite{Soares:2020gqy} one may easily see that
the ultimate effect of a nonvanishing charge is to shrink the overall size
of the event horizon and ergosphere areas. Moreover, taking into account that $\Sigma$ depends on the boost parameter $\gamma$
according to (\ref{kth})--(\ref{sig}), one should expect a deformed ergosphere for a nonvanishing boost parameter.
In Fig. 1 we show 
both effects in a section $\phi={\rm const}$ fixing the parameters $m_0=200$, $\omega_0=195$, $\gamma=0.9$. 
For the boosted Kerr-Newman configurations (red curves) we have considered a charge parameter sufficiently close to its upper limit
which satisfies the cosmic censorship hypothesis, namely $q_0=44$. The reason behind this choice is that configurations which respect the cosmic censorship hypothesis far from these develop an event horizon/esgosphere which coalesce with the pure Kerr boosted case. In the left panel we show the event horizons and ergospheres in Bondi-Sachs coordinates. In the right panel we considered our preferred timelike foliation in the $(t, r, \theta, \phi)$ coordinates.
Apart from the shrunken event horizon and ergosphere, we also see that the developed ergospheres appear deformed in the opposite direction of the boost in both cases, as one should expect.

\section{The Electromagnetic Pattern}

In this section we examine some basic structures of the electromagnetic field when
measured by a locally nonrotating frame of reference, namely a zero angular momentum observer\cite{Takahashi:2007yi}.
To this end, we evaluate the lapse $N$, shift $N^i$
and the spatial metric $\gamma_{ij}$ of (\ref{ks}) considering its ADM decomposition as
\begin{eqnarray}
ds^2=-N^2dt^2+\gamma_{ij}(dx^i-N^idt)(dx^j-N^jdt).    
\end{eqnarray}
\begin{figure*}
\begin{center}
\includegraphics[width=5.5cm,height=5cm]{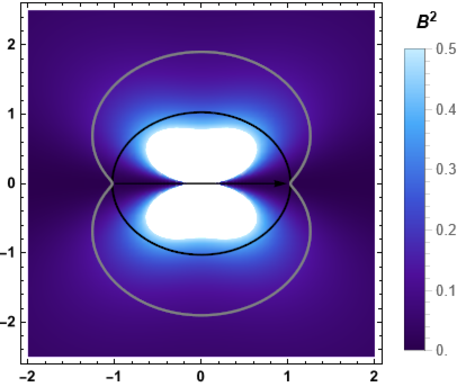}
\includegraphics[width=5.5cm,height=5cm]{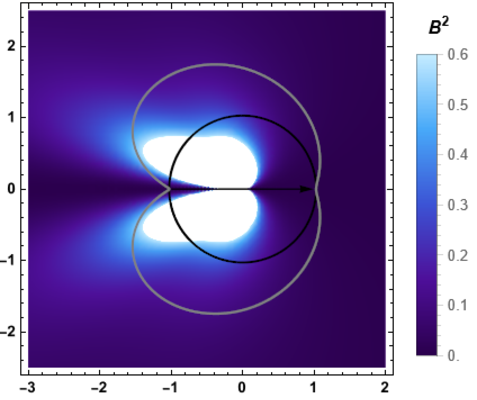}
\includegraphics[width=5.5cm,height=5cm]{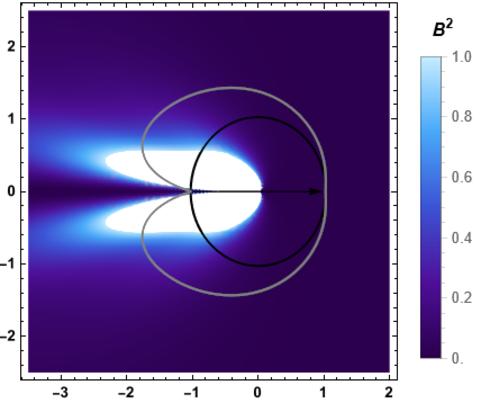}
\caption{The magnitude of the magnetic field for different boost parameters $\gamma=0.00$ (left),
$\gamma=0.75$ (middle) and $\gamma=1.5$ (right). In
these plots the color scales characterize the intensity of the modulus of the respective magnetic fields. For increasing $\gamma$'s the
maximum of the modulus of the intensity appears collimated in the opposite direction of the boost
denoted by the black arrow -- the same direction of the rotation axis. The solid black curves stand for the event horizons while the gray ones characterize the ergospheres.}
\label{fig1}
\end{center}
\end{figure*}
In this case the nonvanishing components of the lapse, shift and spatial metric read
\begin{eqnarray}
&&N=\sqrt{\frac{r^2+\Sigma^2}{r^2+(2m_b r-q^2)K^3+\Sigma^2}},\\
&&N^r=\frac{(2m_br-q_b^2)K^3}{r^2+(2m_br-q_b^2)K^3+\Sigma^2},\\
\label{adm1}
&&\gamma_{rr}=1+\frac{(2m_br-q_b^2)K^3}{r^2+\Sigma^2},\\
&&\gamma_{r\phi}=-\Big[\frac{r^2+(2m_br-q_b^2)K^3+\Sigma^2}{r^2+\Sigma^2}\Big]\frac{\omega_b\sin^2\theta}{K},\\
&&\gamma_{\theta\theta}=\frac{r^2+\Sigma^2}{K^2},\\
\nonumber
&&\gamma_{\phi\phi}=\Big(\frac{r^2+\Sigma^2}{K^2}\Big)\sin^2\theta\\
\label{adm2}
&&~~~~~~+\Big[\frac{r^2+(2m_br-q_b^2)K^3+\Sigma^2}{r^2+\Sigma^2}\Big]\frac{\omega^2_b\sin^4\theta}{K^2}.
\end{eqnarray}
In this context the locally nonrotating frame of reference is defined by the $4$-velocity
\begin{eqnarray}
u^\mu=\frac{1}{N}(1, N^r, 0, 0).    
\end{eqnarray}

By defining the Hodge dual ${\cal F}^{\mu\nu}=\frac{1}{2}\epsilon^{\mu\nu\alpha\beta}F_{\alpha\beta}$
we are able to evaluate the contravariant components of the electric and magnetic field
as
\begin{eqnarray}
E^\mu=F^\mu_{~\nu}u^\nu,~~B^\mu={\cal F}^\mu_{~\nu}u^\nu.    
\end{eqnarray}
It is then easy to show that the only nonvanishing component 
-- up to second order -- of the electric field reads 
\begin{eqnarray}
\label{er}
E^r\simeq\frac{\sqrt{2} }{\kappa}\Big(\frac{q_0}{r^2 }\Big).    
\end{eqnarray}
On the other hand, the magnetic counterpart read
\begin{eqnarray}
\label{br}
B^r&\simeq&-\frac{2\sqrt{2}q_0}{\kappa r }\Big[\frac{\Sigma(\theta)\sin\theta}{ K^2(\theta)}\Big]\Big(1-\frac{m_0}{r}\Big),\\
\label{bt}
B^\theta&\simeq&-\frac{\sqrt{2}q_0\omega_0\sin^2\theta}{\kappa r^2 K^2(\theta)}.
\end{eqnarray}
%
Comparing (\ref{er})--(\ref{bt}) with the results obtained in \cite{Aranha:2021zwf} we see that apart from a radial electric field which falls
with $r^{-2}$, there is an additional component of the magnetic field, namely $B^\theta$, which
provides a rather involved structure of the magnetic field due to the boost parameter.

To better understand the physics behind such structure we consider the magnitude of the magnetic field. Taking into account the spatial part of the metric according to the ADM decomposition(\ref{adm1})--(\ref{adm2}), straightforward calculations furnish 
\begin{eqnarray}
E^2 \simeq \mathcal{O}(1/r^4)    
\end{eqnarray}
while
\begin{eqnarray}
B^2\simeq 2\Big(\frac{q_0\sin\theta}{\kappa r K^2
(\theta)}\Big)^2\Big[\frac{\omega_0^2\sin^2\theta}{K^2(\theta)}+4\Sigma^2(\theta)\Big]    
\end{eqnarray}
up to second order in $1/r$. Once the corrections in (\ref{br})--(\ref{bt}) are of 
second order in $1/r$, we expect to obtain an analogous behaviour of the magnetic field 
when compared to \cite{Aranha:2021zwf}. That is, two intense lobes of the magnetic field should be
observed in the direction opposite to that of the boost. To illustrate this behaviour
we fix $\kappa=1$ and the parameters $m_0=1$, $\omega_0=0.9$, $q_0=0.435$.
A careful reader may notice that these parameters are connected to quasi-extremal configurations.
The reason behind this choice is that configurations which respect the cosmic censorship hypothesis far from these develop a weak magnetic field beyond the event horizon region so that the 
contribution to the expectable lobes appear negligible. In Fig. 2 we show the behavior of $B^2$ for
different boost parameters.

\section{Final Remarks}

In this paper we obtain an approximate black hole solution of Einstein field equations which describes a boosted Kerr-Newman black hole relative to a Lorentz frame at future null infinity. 
Assuming a nonvanishing electric charge and that the boost is aligned with the black hole angular momentum we obtain a solution of the field equations up to $1/r^4$ in a preferred timelike foliation. We show that the
ultimate effect of a nonvanishing charge is to shrink the overall size of the event horizon and ergosphere areas when compared to the neutral boosted Kerr black holes. 
For a proper nonrotating frame of reference we evaluated the electromagnetic fields. 
We show that while the electric field displays a pure radial
behaviour, the magnetic counterpart develops an involved structure with 
two intense lobes of the magnetic field observed in the direction opposite to the boost.

As a future perspective we intend to examine electromagnetic aspects considering different astrophysical configurations.
In particular, the behaviour of accelerated test particles in the background described by (\ref{ks}) should be of physical interest
in the context of Blandford-Znajek processes\cite{Blandford:1977ds}. Numerical results 
on plasma simulations of black hole jets\cite{Parfrey:2018dnc} should also be compared with our method.
For the last but not least, we also intend to generalize our method considering an arbitrary boost direction.
To this end we shall extend the method shown in this paper considering the neutral case\cite{Aranha:2022fuq} as a particular configuration.

\section{Acknowledgments}

RM acknowledges financial support from
FAPERJ Grant No. E-$26/010.002481/2019$.


\end{document}